\documentclass[a4paper]{article}
\usepackage{RR-Inria/RR}
\usepackage{hyperref}
\usepackage{algorithm2e}
\usepackage{graphicx}
\usepackage{amsfonts}
\usepackage{epsf}\epsfverbosetrue
\usepackage{alltt}
\usepackage{subfigure}
\usepackage{color}
\usepackage{float}
\usepackage{url}

\newcommand{\HRule}{\rule{\linewidth}{0.5mm}}

\usepackage[T1]{fontenc}
\usepackage{ulem}
\usepackage[final]{changes} 
\usepackage{amsmath,latexsym}
\usepackage{times}
\usepackage{cite}

\usepackage{graphicx,epsfig} 
\usepackage{listings}
\usepackage{color}
\usepackage{epsf}\epsfverbosetrue
\usepackage{alltt}
\usepackage{subfigure}

\usepackage{algpseudocode}
\usepackage{ioa_code}

\usepackage{boxedminipage, epsfig,graphics,graphicx,subfigure, amsmath,amssymb,multirow, array}
\usepackage{newalg}
\usepackage{url}
\usepackage{color}

\def\BibTeX{{\rm B\kern-.05em{\sc i\kern-.025em b}\kern-.08em
    T\kern-.1667em\lower.7ex\hbox{E}\kern-.125emX}}

\newcommand{\remove}[1]{}

\newcommand{\psfigure}[3]{ 
 \begin{figure}[\placement]\begin{center}%
  \epsfig{file=figs/#2,width=#1\hsize}%
\let\normalsize\small\caption{#3\label{fig:#2}}%
  \end{center}
\end{figure}}

\definechangesauthor{FLF}{blue}
\definechangesauthor{AV}{green}
\definechangesauthor{CS}{red}
\definechangesauthor{AP}{magenta}

\RRdate{December 2009} 
\RRNo{7288}

\def\email#1{~}

\RRauthor{
 Mubashir Husain Rehmani\thanks{Lip6/Universite Pierre et Marie Curie - Paris 6, France} 
 \and Aline Carneiro Viana\thanks{INRIA, France and TU-Berlin, Germany}
 \and Hicham Khalife\thanks{LaBRI/ENSEIRB, France}
 \and Serge Fdida\thanks{Lip6/Universite Pierre et Marie Curie - Paris 6, France}
 }

\authorhead{Husain Rehmani \& Carneiro Viana \& Khalife \& Fdida}
\RRtitle{Dissémination fiable de données dans les réseaux radio cognitifs multi-sauts}
\RRetitle{Toward Reliable Contention-aware Data Dissemination in Multi-hop Cognitive Radio Ad Hoc Networks} 
\titlehead{Toward Reliable Contention-aware Data Dissemination in Multi-hop CRAHNs}
\RRresume{Nous présentons dans cet article une nouvelle stratégie de sélection de fréquences pour la dissémination fiable des données dans le réseau radio cognitifs multi-sauts. Le principal défi ici est de sélectionner des fréquences offrant un bon compromis entre la connectivité et les contentions. En d'autres termes, il s'agit de sélectionner les fréquences offrant de bonnes opportunités de communication en raison de (1) la faible activité des n\oe{}uds radio primaire (PR), et (2) la contention limitée entre les n\oe{}uds ratios cognitifs (CR) opérant sur ces fréquence. Ainsi, en explorant dynamiquement les ressources résiduelles sur les fréquences des primaires et en contrôlant le nombre de CR sur une fréquence particulière, notre stratégie SURF permet de construire un réseau connecté à contention limitée où des communications fiables peuvent être effectuées. Grâce à une large étude basées sur des simulations, nous étudions les performances de SURF par rapport à trois autres approches. Les résultats de simulation confirment que notre stratégie permet la sélection des meilleures fréquences adaptées à une dissémination fiable et efficace des données dans un réseau radio cognitifs multi-sauts.

}
\RRabstract{
\begin{abstract}

This paper introduces a new channel selection strategy for reliable contention-aware data dissemination in
multi-hop cognitive radio network.  The key challenge here is to select channels providing a good tradeoff
between connectivity and contention. In other words, channels with good opportunities for communication due
to (1) low primary radio nodes (PRs) activities, and (2) limited contention of cognitive ratio nodes (CRs)
acceding that channel, have to be selected. Thus, by dynamically exploring residual resources on channels and
by monitoring the number of CRs on a particular channel, SURF allows building a connected network with
limited contention where reliable communication can take place. Through simulations, we study the performance
of SURF when compared with three other related approaches. Simulation results confirm that our approach is
effective in selecting the best channels for efficient and reliable multi-hop data dissemination.
\end{abstract}
}
\RRmotcle{Réseaux radio cognitifs multi-sauts, routage opportuniste, sélection dynamique de fréquence, dissémination de données}

\RRkeyword{multi-hop cognitive radio networks, opportunistic forwarding, dynamic channel selection, data dissemination.} 
\RRprojets{Asap}
\RRtheme{\THCom}
\RCSaclay 

\begin{document}
\normalem
\makeRR   

\def\baselinestretch{1.5} 

\section{Introduction}
\label{sec:in}

\def\baselinestretch{1.5} 

Data dissemination is a classical and a fundamental function in any kind of network. In wireless
networks, the characteristics and problems intrinsic to the wireless links bring several challenges in data
dissemination in the shape of message losses, collisions, and broadcast storm problem, just to name a few.
However, in the context of Cognitive Radio Network (CRN)~\cite{survey}, reliable data dissemination is much
more challenging than traditional wireless networks. 
First, in addition to the already known issues of wireless environments, 
the diversity in the number of channels each cognitive node can use adds another
challenge by limiting node's accessibility to its neighbors. Second, cognitive
radio nodes have to compete with the Primary Radio (PR) nodes for the residual resources on many channels and
use them opportunistically. Besides, during communication CR nodes should communicate in such a way that
it should not degrade the reception quality of PR nodes by causing CR-to-PR interference. In addition, CR
nodes should immediately interrupt its transmission whenever a neighboring PR activity is
detected~\cite{hicham}. 

\begin{figure}[h]
    \begin{center}
    \includegraphics[width=9.5cm]{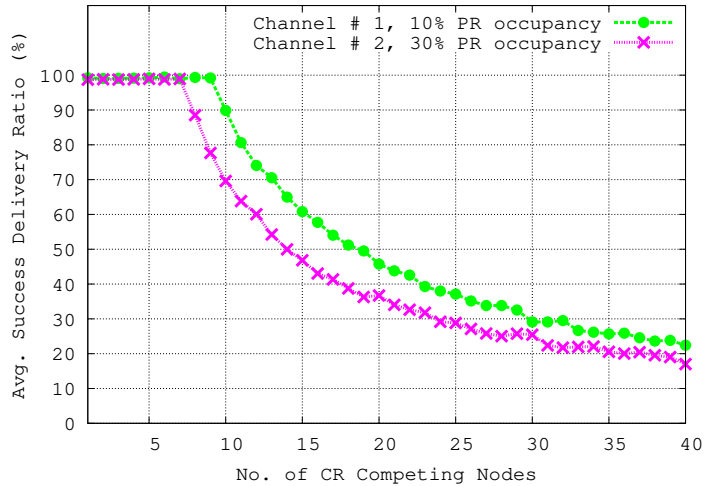}
\caption{CR nodes competing for the same channel.}
    \label{fig1-a}
\end{center}
\end{figure}

In multi-hop cognitive radio ad hoc networks, where coordination between CRs is hard to achieve and 
no central entity for regulating the access over channels is to be envisaged,
reliable data dissemination is even more complex. 
In this perspective, the first step in
having efficient data dissemination is  to know {\it how to select best channels}. Thus, differently from
most works in the literature dealing with single-hop communication~\cite{cao,jcm}, we go a step further
here and build up a channel selection strategy for multi-hop communication in CRN.
The objective of every cognitive radio node is to select the best channel ensuring a maximum
connectivity and consequently, allowing the largest data dissemination in network. This corresponds to
the use of channels having not only low primary radio nodes (PRs) activities, nevertheless the reliability 
of the dissemination process is achieved by limiting the contention of
cognitive ratio nodes (CRs) acceding selected channels.

The effect of CR contentions on dissemination is highlighted in Fig.~\ref{fig1-a} that shows the evolution of the average 
success delivery ratio at receivers of a single
source, with the number of competing CRs. 
It is clear that the performance of a channel with low PR activity
decreases with the number of CR competing for the available resource. Nevertheless, a channel with higher PR activity can be a
good choice if CR contention is low. The challenge here is then how to find a good tradeoff
between connectivity and contention.

In this paper, we propose a channel selection strategy, named SURF. The goal of SURF is to ensure reliable
contention-aware data dissemination and is specifically designed for multi-hop cognitive radio ad hoc
networks. Usually channel selection strategies provide a way to nodes to select channels for transmission.
On the contrary, SURF endue CR nodes to select best channels not only for transmission but also for
overhearing. As a result, both sender and receiver tuned to the right channel for effective and reliable data
dissemination. Additionally, by dynamically exploring residual resources on channels and by monitoring the
number of CRs on a particular channel, SURF allows building a connected network with limited contention where
reliable communication can take place. To counter this issue, we define the ``Tenancy Factor $\beta$'',
which enables SURF algorithm to avoid channels with high CR contention.

Through simulations, we show that SURF builds, as expected, a highly connected network 
suitable for reliable dissemination. 
Moreover, SURF outperforms existing algorithms. 
In fact, we compared our solution with two variants of Selective Broadcasting (SB) strategy~\cite{agrawal},
the closest technique to SURF available today.
The simplicity and decentralized nature of our 
solution makes it usable in ad hoc CRNs deployed to convey emergency messages and alerts. It can also be employed in
commercial applications to disseminate short publicity messages.


The remainder of this paper is organized as follows: we discuss connectivity vs.
contention trade-off in Section~\ref{motiv}. We give general overview of SURF in Section~\ref{overview}.
Section~\ref{sec:proposal} deals with detailed description of SURF. Section~\ref{imp_para} provides comprehensive analysis of SURF. Performance analysis is done in
section~\ref{sec:analysis}, then the major advantages of SURF are highlighted in section~\ref{adv}. Section~\ref{related} discuss related work and finally,
section~\ref{sec:conclusion} concludes the paper.

\section{Cognitive radio ad hoc networks: connectivity VS. contention trade-off}
\label{motiv}

In a highly dynamic/opportunistic cognitive radio network, cognitive users compete for residual 
resources (a.k.a spectrum holes) left by the activity
of the legacy users more formally called primary radio users. Every cognitive node, using an intelligent selection strategy, selects the appropriate
channel for transmitting with the major constraint of not degrading the service of ongoing primary radio communications. Indeed,
primary radios have the absolute priority over the communication channels.
In an opportunistic multi-hop cognitive radio network where coordination between CRs is hard to achieve and no central entity for regulating
the access over channels is to be envisaged, the objective of every cognitive radio is to select the channel ensuring a maximum connectivity.
Such spectrum band has the highest number of active cognitive radios hence allows quick and effective data dissemination in the network.

\begin{figure}[h]
\begin{center}
    \includegraphics[width=12cm]{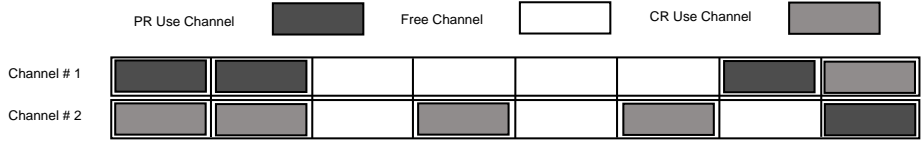}
\caption{PR and CR Nodes occupancy over channels}
    \label{fig1-b}
    \end{center}
\end{figure}

Intuitively, one may think that the best strategy for \emph{all} CRs is to dynamically switch to the less
occupied channel (by PRs). Thus, satisfying the objective of verifying priority constraints imposed by PRs.
Nevertheless, such strategy leads to
many classical problems already well known in wireless networking. First, forcing all CRs in a geographic
area to be active over the same channel makes all nodes compete for the same resource thus generating
contention and collision problems. Second, such approach wastes the valuable additional capacity on different
channels that the cognitive radio concept offers. Indeed, it was already shown in traditional wireless
networking that networks with high contention, where repetitive collisions are frequent, suffer from
close to zero throughput~\cite{jingyang}. A typical example is described in Fig.~\ref{fig1-b}.
Initially, channel 1 has more primary radio activities and should be avoided by the CR transmitters. However,
if enough CRs switch to channel 2 to communicate, channel 1 quickly becomes less occupied and able to carry
higher throughputs than channel 2. Therefore, taking into account contention issues due to CR transmissions is
necessary when selecting spectrum bands for CR communications.

Any proposed strategy for channel selection in CRN has to optimize the connectivity vs. contention trade-off.
We propose hereafter a channel selection strategy that monitors the number of active CR nodes on a particular
channel. As a result, {\it we are able to build a well connected network while dynamically exploiting
residual resources on many channels}. We detail how our proposed strategy named SURF handles this trade-off
in the following sections.

\section{SURF: General overview}
\label{overview}

SURF channel selection strategy is specifically designed for ad hoc cognitive radio networks.
The general goal of our strategy is to ensure a reliable data dissemination over a multi-hop CRN. Such technique can
be used to convey emergency alarms and alerts or to deliver low priority data such as advertisement messages 
in a cognitive radio multi-hop
context. Recall that in order to achieve our goal
and ensure coverage and reliability, the connectivity vs. contention trade-off should be optimized.

SURF strategy is exclusively implemented by every CR node and is used for transmission and/or overhearing.
As detailled hereafter, using the decentralized algorithm proposed by SURF, every CR sender judiciously selects the \textit{best} 
frequency band for sending messages and every CR receiver tunes to the right channel (selected by the sender) 
to retrieve the sent data.

With SURF, each CR node looks first for the less PR-occupied 
channel to help deciding autonomously which channel to use.
In addition to PR occupancy, we also consider CR neighbors competing for the same channel resource.
More precisely, every CR node classifies available channels based on the observed PR-occupancy over these 
channels. This classification is then refined by identifying the number of active CRs over each band. The
best channel for transmission is the channel that has the lowest PR activity and a reasonable ongoing CR activity.
Indeed, choosing a channel with few CRs yields to a disconnected network. The challenge in our strategy is in 
finding the number of active CRs on every channel that gives the best connectivity with limited contention.
Practically, every CR after classifying available channels, switches dynamically to the best one and broadcasts the stored message.

Additionally, CRs with no messages to transmit implement the SURF strategy in order to tune to
the \emph{best} channel for data reception. Clearly, using the same strategy implemented by the sender allows
nodes in the close geographic areas to select the same channel as sender for overhearing with high probability. 
Intuitively, it is likely that CRs in the sender's vicinity have the same PR occupancy, hence channels available 
to a CR sender is also available to its neighbors with high probability~\cite{pomdp}.
Therefore, SURF controls the number of CR receivers, thus a connected topology with low contention is created.
Once a packet is received, every CR receiver undergoes again the same procedure to choose the appropriate
channel for conveying the message for its neighbor.

\section{Detailed description of SURF}
\label{sec:proposal}
\subsection{Considered Scenario}
\label{scenario}

We consider an infrastructureless multi-hop cognitive radio ad hoc network in which only CR nodes collaborate.
In the assumed configuration, no cooperation or feedback from primary nodes is expected.
Consequently, CR nodes can only rely on information obtained or inferred locally to undergo transmission or reception decisions.

Moreover, we consider that CR nodes are capable to switch over the available channels easily. For the sake of
simplicity, we assume that every channel is divided into equal time slots. Each slot can be exclusively used
either by a primary radio or by a cognitive radio when no primary activity is present over that slot. Hence,
every channel is composed of $\tau_t=\tau_o + \tau_a$ time slots, where $\tau_o$ and $\tau_a$ are the slots
occupied by PR nodes and the available slots, respectively. A total of $C$ frequency channels are available
in the network. These channels can either be used by PR or CR nodes. Due to PR nodes' localized and timely
activity, the total frequency channels vary with time and location, which results in a non-uniform, scattered
and diverse set of channels available to CR nodes. 

We consider CR nodes equipped with a single transceiver, where a single channel can be selected at a 
time and used \emph{exclusively} for transmission or overhearing. It is worth noting here that, contrarily to other
approaches in the literature where the costly assumption of having CR nodes equipped with multi-transceivers
is used, {\it our assumption is highly realistic today} and is already considered in some cognitive radio devices
and prototypes~\cite{bib:harada_pimrc2008}. In these devices, physical constraints limit the access of CR
nodes to a limited set of available channels. Indeed, covering all the spectrum bands is a highly costly
process, thus we assume that every cognitive radio device can handle a predefined number of channels. We
denote the set of spectrum channels each CR can exploit by $Acs$ such that $Acs \in C$ and $|Acs| < |C|$. We
shall investigate the impact of the size of $Acs$ on the performance of our strategy later in the paper
(Section \ref{Acs}). In the considered scenario described above, SURF strategy provides every CR with a way to
select the appropriate channel for transmission and reception.

\subsection{PR and CR Occupancy}
\label{Proccup}

We consider that the spectrum sensing block provides the spectrum opportunity map as described in ~\cite{arslan}, which is
then used by CR nodes to calculate the PR and CR occupancy. The spectrum sensing block is responsible for
obtaining awareness about the spectrum usage and presence of primary users; whereas spectrum opportunity map
identifies whether primary users have been detected or not for each channel. PR occupancy is thus, denoted here by
$PR_o^{(i)}$ and is defined as the time slots percentage of the channel $i$ occupied by PR nodes, i.e. the ratio
between the number of PR nodes and the total number of time slots $\tau_t$.

Practically, our estimation of the PR occupancy follows a conservative approach. In fact, computing the PR
occupancy based on the total number of primary radios assumes these nodes are permanently active.
Intuitively, given the higher priority of PRs in accessing spectrum bands, considering that PR are always
active gives them additional guarantees at the price of a lesser space for CRs activity. The remaining
available percentage of the channel $i$, i.e. $1 - PR_o^{(i)}$, gives then the space available for channel
sharing among CR nodes, named $CR_{as}^{(i)}$. The CR occupancy $CR_o^{(i)}$ is then obtained from the
available space for CR activities on a particular channel $i$, $CR_{as}^{(i)}$, as described hereafter.

Since we are considering non-cooperative infrastructureless architecture, there is no centralized authority
that helps CR nodes for their channel selection. Therefore, there is no way to prevent collision and message
losses when the number of CR nodes competing for the same channel increases. Additionally, CR nodes have to
rely on {\it locally} inferred information in a distributed manner to select channels with a higher number of
1-hop CR receivers. To allow nodes to select channel having a good compromise between the number of CR
receivers and the number of competing CR transmitters, we use the {\it Tenancy Factor}, named $\beta$,
to compute the CR occupancy $CR_o^{(i)}$ of each channel $i$. $\beta$ provides the upper bound in terms of
number of CR neighbors on a particular channel, where the communication is still performed with a good
probability of success. The goal here is then to maximize the chances of selecting channels that have a
a good number of CR neighbors (close to {\it Tenancy Factor} $\beta$).

\begin{figure}[h]
    \HRule \\[0.4cm]
\footnotesize
\textbf{if} $CR_n^{(i)} \textless  \beta$ \\
\hspace{0.5cm}    \textbf{then}  $ CR_o^{(i)} \leftarrow \frac{CR_{as}^{(i)}}{(\beta - CR_n^{(i)})}$\\
\textbf{else if} $CR_n^{(i)} = \beta$ \\
\hspace{0.5cm}    \textbf{then} $ CR_o^{(i)} \leftarrow CR_{as}^{(i)} $ \\
\textbf{else} $CR_n^{(i)} \textgreater \beta$ \\
\hspace{0.5cm}    \textbf{then}  $ CR_o^{(i)} \leftarrow \frac{CR_{as}^{(i)}}{CR_n^{(i)}}$\\
\textbf{end if}\\   
\hspace{0.5cm}    \HRule \\[0.4cm]
\caption{Algorithm for CR occupancy's computation}
    \label{Alg}
\end{figure}


Figure~\ref{Alg} shows algorithm how
CR nodes calculate CR occupancy according to the tenancy factor $\beta$ and the number of CR neighbors
competing for the channel. When the number of CR neighbors, i.e.  $CR_n^{(i)}$, on a particular 
channel $i$ is lower than $\beta$, the chances of the channel with number of neighbors close to $\beta$ 
to be selected increases. The best channel in terms of channel
availability is the one with CR neighbors equal to $\beta$. When CR neighbors are higher than $\beta$, the
higher number of neighbors decreases the chances of the channel to be selected.

\subsection{Channel Selection}
\label{surf}

$SURF$ strategy classifies channels by assigning a weight $P_w^{(i)}$ to each observed channel $i$
in the $Acs$ set. Thus, every cognitive radio running SURF, locally computes the $P_w^{(i)}$ 
using the following equation:

\begin{equation}
\label{pequation} {\forall {\it i} \in {\it C}: P_w^{(i)} = e^{-PR_o^{(i)}} \times CR_o^{(i)} }
\end{equation}

$P_w^{(i)}$ describes the availability level of a channel $(i)$ and is calculated based on the occupancy of
PR (i.e. $PR_o^{(i)}$) and CR (i.e. $CR_o^{(i)}$) nodes over this channel (c.f. Section~\ref{Proccup}).
If a CR node finds two or more channels having identical higher values of $P_w$, it firstly tries to
select the one that has lower $PR_o^{(i)}$ among them. If they also have identical values of $PR_o^{(i)}$,
then the CR node randomly selects one channel among the channels with identical and higher values of $P_w$.

Practically, the computed
availability in Eq.~\ref{pequation} exponentially decreases with the PR occupancy and linearly increases
with the available space for CR activities. These two behaviors are directly related to the two objectives the SURF
strategy needs to satisfy. The major objective of protecting the ongoing PR activity is mapped into an exponential
decrease of a channel weight as a function of the PR occupancy. The higher the PR
occupancy over a spectrum band the exponentially lower the weight will be. Thus, SURF gives high importance
to not degrading the service of ongoing primary communications. The second objective of ensuring a maximum connectivity
is implemented in the second term of Eq.\ref{pequation}. More precisely, the $CR_o^{(i)}$ obtained through Algorithm
(cf. Figure\ref{Alg}) increases with available space for CRs activity, while carefully considering the connectivity versus contention
trade-off.

\section{SURF Comprehensive Analysis}
\label{imp_para}

In this section, we investigate how SURF reacts to different CRN conditions (i.e. PR activity, available channels, etc) 
when different values of $\beta$ are used. The goal is thus, to well understand the $\beta$ effect and to be able to select 
the good one for the SURF performance's evaluation (cf. Section~\ref{sec:analysis}).

\subsection{Methodology}
\label{methodology}

We study the SURF behavior for different values of $\beta$ when varying PR activity, set of available channels 
$Acs$, and total number of channels $C$ in the network. We then evaluate three performance issues: 
(i) the average number of CR neighbors per hop, which is the potential 1-hop receivers at each  
channel of the transmitter's $Acs$ set; (ii) the average number of CR receivers per hop, which is the CRs that correctly received 
the sent packet (i.e., the average number of CR nodes that have selected for overhearing the same channel than the transmitter 
selected for sending and, even if unreliable links are considered, have correctly received the packet), and (iii) 
the average loss ratio per hop, which is the number of lost packets over the total 
number of transmitted packets. Since, the system we explore is highly complex and many parameters can be modified, 
for results tractability and clarity, we modify each of a single parameter while fixing all the others.
The influence of $\beta$ over those performance issues is monitored by first varying the PR occupancy (cf. Section~\ref{fixedpr}), 
second by varying the available channel set (cf. Section~\ref{Acs}), and finally, under a dynamic environment where 
both the available channel set and PR occupancy vary, simultaneously (cf. Section~\ref{deter_beta}).

We run simulations over a cognitive radio specific simulator written in C++.
Within the network, we consider that two nodes can communicate if they use at least one common channel
and if they are within the transmission range of each other. In order to simulate message losses, a
probability $P_s^{(i)}$ of successfully sending a message is assigned to each channel $i$ and equals to:

\begin{equation}
\label{loss}
{P_s^{(i)}} = \dfrac{\tau_a^{(i)}}{CR_n^{(i)}} ~~~~~~~ \forall{ {\it i} \in {\it C} }
\end{equation}

\noindent while loss ratio equals to $1-P_s^{(i)}$. In fact, equation (\ref{loss}) states that the
probability of sending a message is dependent on the available slots $\tau_a$ and the number of CR nodes
competing for the channel i.e. $CR_n^{(i)}$. We consider $1\%$ of message losses, if $CR_n^{(i)} \, \textless
\, \tau_a$. It is worth noting here that no retransmission in implemented in our simulations. 

Results are generated from an average of 1000 simulations, along
with 95\% of confidence intervals. For all results, unless otherwise specified, we consider 30 PR nodes and $\tau_t$=6 total 
time slots for each channel. To ensure total network connectivity, the transmission range is set to {\it R} = 250m and
the average CR neighbor density $d_{avg}$ before $Acs$ computation, is set to 20.
We consider $Acs$ size of [5,4,3] for 5 total number of channels ($Ch=5$), 
and [15,12,8] for 15 total channels ($Ch=15$). 
The number of CR nodes is fixed to {\it N=70} and randomly deployed within a square area of {\it $a^2=$}707x707$m^{2}$~\cite{gupta}.
Because the total number of CRs is fixed, it is straightforward to notice that the number of average CR neighbors per 
channel decrease as the $Acs$ size decrease. 
{\it TTL} is introduced to disseminate the message in the whole network. It is the maximum number of hops
required for a packet to traverse the whole network and is set to $\lceil \dfrac{2a}{R} \rceil$, i.e. $TTL=6$ in our simulation
scenario. The total number of PRs are uniformly distributed among the existing channels.

\subsection{Impact of PR Occupancy}
\label{fixedpr}


\begin{figure}[h]
   \begin{center}
   \includegraphics[width=10cm,height=7cm]{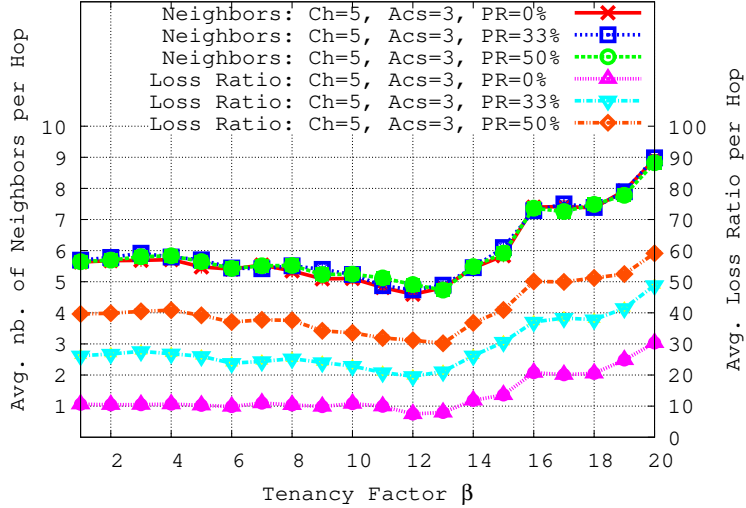}
 \caption{Tenancy factor $\beta$, average number of neighbors per hop, and average loss ratio per hop, in a CRN with 70 CR nodes for varying PR occupancy (i.e. fixed time slots occupied by PR nodes) and fixed $Acs$, for channels=5.}
   \label{fig2a}
\end{center}
\end{figure}


\begin{figure}[h]
   \begin{center}
   \includegraphics[width=10cm,height=7cm]{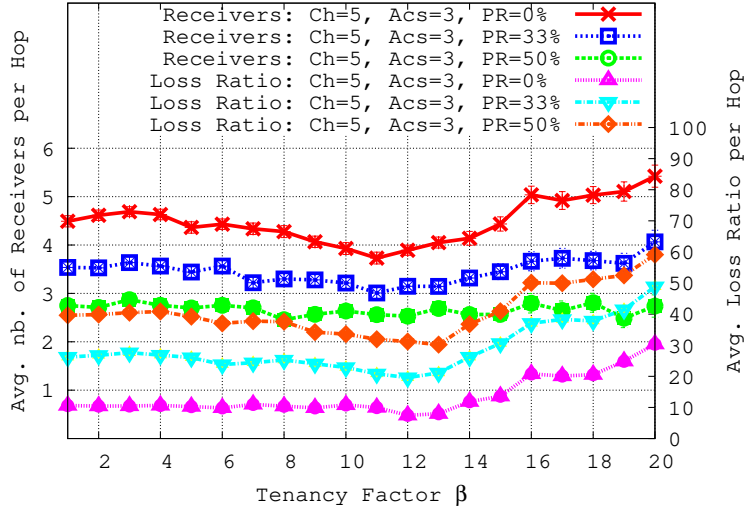}
 \caption{Tenancy factor $\beta$, average number of receivers per hop and average loss ratio per hop, in a CRN with 70 CR nodes for varying PR occupancy (i.e. fixed time slots occupied by PR nodes) and fixed $Acs$, for channels=5.}
   \label{fig2}
\end{center}
\end{figure}

Higher (cf. lower) PR occupied channels give lower (cf. higher) space for CR communication. Thus, to investigate the impact of 
PR occupancy over the analysed performance issues, we varied the PR occupancy from 0\%, 33\%, and 50\% by changing the 
slots occupied by PR nodes to 0, 2, and 3, respectively. We then fixed the total number of channels $Ch$ to 5 and the 
size of the $Acs$ set per node to 3, which results in an average density of approximately 8 CR neighbors per 
channel of the $Acs$ set. 

Fig.~\ref{fig2a} and Fig.~\ref{fig2} shows then the average number of CR neighbors and 
CR receivers per hop (left axes), and the loss ratio per hop (right axes) for varying values of $\beta$.
In particular, those figures show how the PR occupancy impacts the CR contention and consequently, the loss ratio  
when different values of $\beta$ are used. The number of neighbors (cf. Fig.~\ref{fig2a}) 
remains constant for values of $\beta$ lower than 8, 
i.e. when $CR_n^{(i)} \geq \beta$. In this case and according to the SURF algorithm (cf. Figure~\ref{Alg}), 
low values of $\beta$ do not play any role in limiting network contention among CR nodes. Thus, channels are being weighted 
based only on the total number of available slots (i.e. $CR_{as}^{(i)}$) 
or CR neighbors competing for the channels (i.e. $CR_o^{(i)} = \frac{CR_{as}^{(i)}}{CR_n^{(i)}}$). 
On the other hand, the increase of $\beta$ for values higher than $CR_n^{(i)}$, increases the chances of selecting channels 
that have higher number of neighbors, and consequently, higher number of receivers. This happens until a certain 
threshold, when the channel contention is increased and higher loss ratio is detected, affecting the 
number of receivers. This can be perceived for values of $\beta$ higher than 8, as shown in Fig.~\ref{fig2}. 
Note that, since no retransmission of lost messages is implemented, the contention among nodes is limited to their
first try for transmitting a message. The same happens for loss ratio measurement, which is also limited to 
losses at the first transmission of each message only. This can explain why the increase in loss ratio doe not
significantly decrease the number of receivers in Fig.~\ref{fig2}.

Additionally, as expected, with the increase of PR occupancy, the average number of CR receivers per hop decreases and 
the average loss ratio per hop increases. This is primarily because of the lower available slots $CR_{as}^{(i)}$ to 
CR nodes communicating, which leads to more contention and collisions. 
Particularly, when there is no PR occupancy, i.e. PR occupancy=0\%, the average number of receivers per hop is 
the highest and the average loss ratio per hop is the lowest (cf. Fig.~\ref{fig2}), for any value of $\beta$. 
This is due to the fact that 0\% of PR occupancy 
yields to higher available channel slots $\tau_a$, which results in less CR contention. On the other hand, 
as the number of available channel slots $\tau_a$ decreases (e.g. PR occupancy=50\%), CR nodes finds less opportunity
to communicate, resulting in higher loss ratio due to higher contention.

\subsection{Impact of Available Channel Set $Acs$}
\label{Acs}


\begin{figure}[h]
   \begin{center}
   \includegraphics[width=10cm,height=7cm]{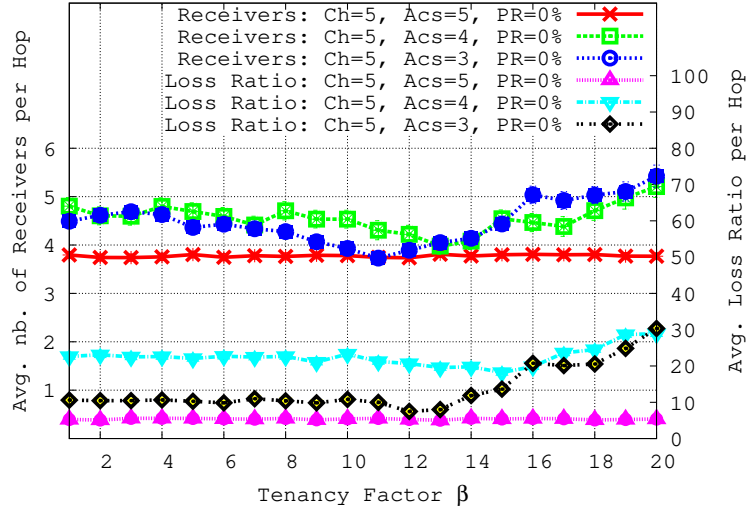}
 \caption{Tenancy factor $\beta$, average number of receivers per hop and average loss ratio per hop, in a CRN with 70 CR nodes for varying number of Acs and fixed PR occupancy, for channels=5.}
   \label{fig4}
\end{center}
\end{figure}


\begin{figure}[h]
   \begin{center}
   \includegraphics[width=10cm,height=7cm]{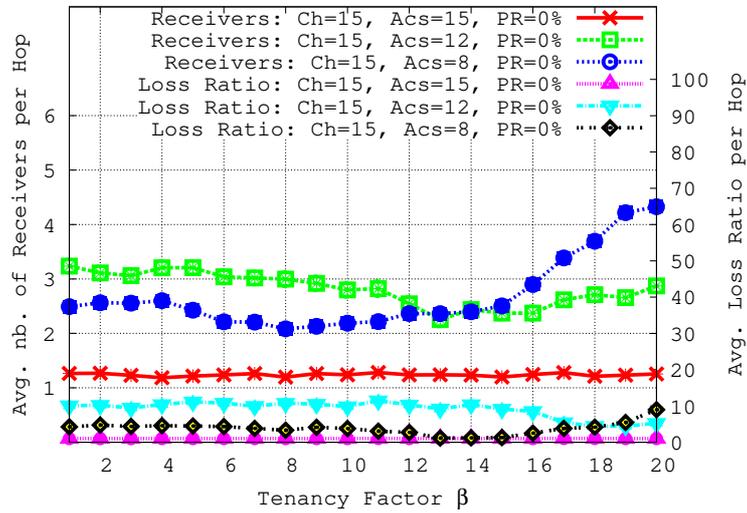}
 \caption{Tenancy factor $\beta$, average number of receivers per hop and average loss ratio per hop, in a CRN with 70 CR nodes for varying number of Acs and fixed PR occupancy, for channels=15.}
   \label{fig5}
\end{center}
\end{figure}

The size of the $Acs$ set and the diversity in number of channels each cognitive node can use, limit node's accessibility 
to its neighbors. Thus, if the $Acs$ set is too small compared to total channels $C$ and diverse for each CR, the number of 
receivers is reduced. In Fig.~\ref{fig4} and Fig.~\ref{fig5}, we evaluate this effect for total number of channels equal to
$Ch=5$ and $Ch=15$, respectively, and, for clarity reasons, consider that all channels are 
unoccupied by PR nodes (i.e. PR occupancy=0\%). We vary the size of the $Acs$ set to 4 and 3 for $Ch=5$ and to 12 and 8 for $Ch=15$.

It can be verified that when a node is accessible by its neighbors on all the channels, $Ch=Acs$ (i.e. all CRs can overhear 
all channels), $\beta$ does not impact on the average number of receivers and loss ratio. This is due to the fact that all the channels 
will have the same computed weight $P_w^{(i)}$: the same PR occupancy (PR occupancy=0\%) and number of CR neighbors 
$CR_n^{(i)}$ is perceived on all channels.
Thus, CR nodes randomly select the channel for transmission and/or 
overhearing. Indeed, a random selection of a channel for transmitting and overhearing
reduces the chances of message reception by neighboring CR. Moreover, this effect is even aggravated by the increase of the 
number of available channels for communication, note the decrease from 4 receivers to 1 for $Ch=5$ and $Ch=15$, respectively 
(cf. Fig.~\ref{fig4} and Fig.~\ref{fig5}).

The decrease of the $Acs$ set's size to 4 and 3 for $Ch=5$ and to 12 and 8 for $Ch=15$, imposes some diversity 
at the number of neighbors per channel, resulting in different weights $P_w^{(i)}$ being assigned to channels. In particular,
the average density of CR nodes per channel is: 12 for $Acs=4$ and for $Acs=12$ and 8 for $Acs=3$ and for $Acs=8$.
That neighborhood diversity implies different levels of contention per channel, which consequently, makes the use 
of $\beta$ impacting the average number of receivers and loss ratio, since channels will be assigned to varying weight values. 
It can be seen from Fig.~\ref{fig4} and Fig.~\ref{fig5} that for lower values of $\beta$ (i.e. when $CR_n^{(i)} \geq \beta$),
CR nodes try to 
select those channels that have number of neighbors close to $\beta$. In this case, the average number of receivers and the average 
loss ratio is lower due to less contending nodes for same channel resource. Whereas with the increase of 
$\beta$ to values higher than  $CR_n^{(i)}$ (i.e. higher than 12 for $Acs=4$ and $Acs=12$ or 8 for $Acs=3$ and $Acs=8$), 
we notice an increase in the average number of receivers as well as the average loss ratio. As previously mentioned, this happens
due to the fact that message retransmission is not implemented in our simulations. Thus, the contention increase caused by the
increase of $\beta$ does not significantly affects the average number of receivers.

\subsection{Choosing the correct value of Tenancy Factor $\beta$}
\label{deter_beta}


\begin{figure}[h]
   \begin{center}
   \includegraphics[width=10cm,height=7cm]{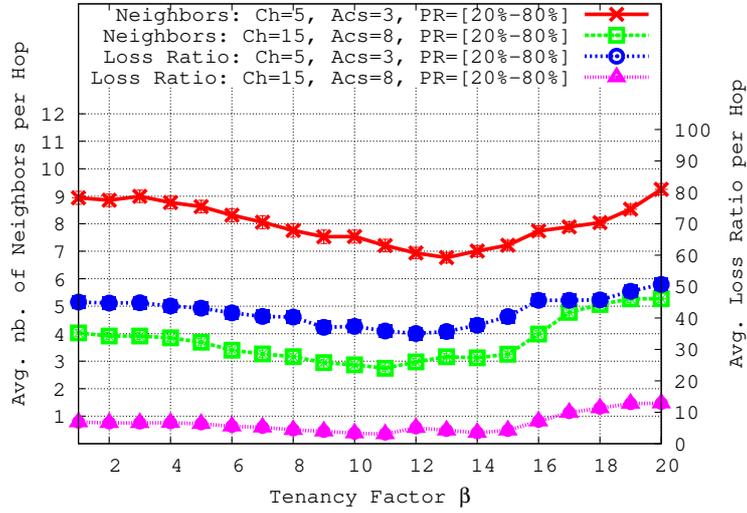}
 \caption{Tenancy factor $\beta$, average number of neighbors per hop, and average loss ratio per hop, in a CRN with 70 CR nodes for varying PR occupancy in the range [20\%-80\%] by PR nodes.}
   \label{fig13}
\end{center}
\end{figure}


\begin{figure}[h]
   \begin{center}
   \includegraphics[width=10cm,height=7cm]{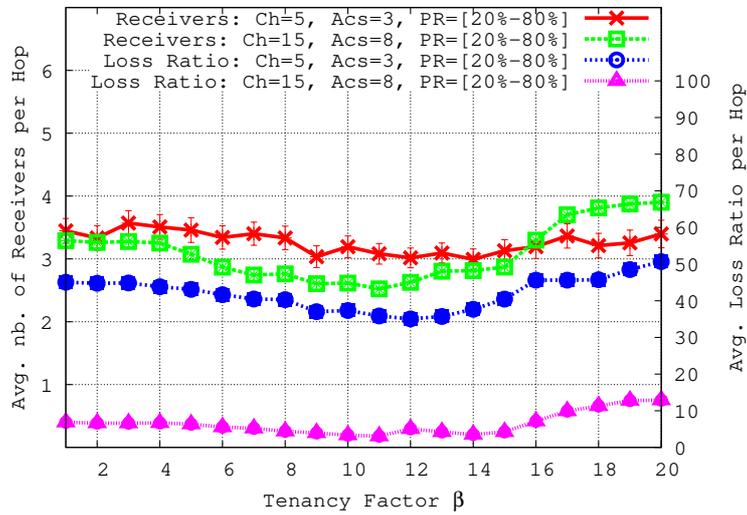}
 \caption{Tenancy factor $\beta$, average number of receivers per hop and average loss ratio per hop, in a CRN with 70 CR nodes for varying PR occupancy in the range [20\%-80\%] by PR nodes.}
   \label{fig6}
\end{center}
\end{figure}

In real environments, channels can be available in some parts of the network and occupied in others. 
Thus, to incorporate this notion, we consider varying PR occupancy and limited available channels. We then set PR occupancy to the 
range of [20\%-80\%] for each PR node over each channel and $Acs$ size to 3 and 8 for $Ch=5$ and $Ch=15$, respectively. 
Here, we investigate how SURF adapts under this dynamic environment with varying PR occupancy and limited $Acs$ sets. 
We perform experiments in order to determine appropriate value of $\beta$ to be used in SURF performance analysis presented
in the next section.

In Fig.~\ref{fig13} and Fig.~\ref{fig6}, we investigate the impact of $\beta$ on average number of neighbors, number of receivers,
and loss ratio per hop. Clearly, the best value of $\beta$ is the one that provides a good tradeoff between number of receivers and 
loss ratio. In this case, for $Ch=5$, the best value of $\beta$ is $\beta=10$ (cf. Fig.~\ref{fig6}). For $Ch=15$, 
the low average number of 5 neighbors on channels (cf. Fig.~\ref{fig13}) is not enough to cause high contention and 
consequently, to decrease the receiver number when high values of $\beta$ are used, as shown in Fig. ~\ref{fig6}. 
Therefore, it is better to use the channels with higher number of neighbors, and thus, a higher value of beta. 
For this reason, we select $\beta=18$ for $Ch=15$, as it provides a good tradeoff between receivers and loss ratio.

Note that, in our simulations, we consider that the same number of PRs is spread over the available channels. 
Thus, using more channels reduces the number of PRs over each channel and consequently reduces the PR occupancy, resulting in 
higher available space for CR nodes for communicating. Therefore, the average loss ratio per hop for $Ch=15$ is much lower than 
$Ch=5$. This is due to the fact that CR nodes find more space and hence, causes less contention for the same channel resource. 
On the other hand, the increase of the number of channels increases the probability of having
two or more channels assigned to the same $P_w^{(i)}$ value, increasing consequently the probability of 
having a random selection of channels.

\section{Performance Evaluation}
\label{sec:analysis}

\subsection{Simulation Environment}
\label{surfanalysis}

In order to evaluate the performance of SURF, we compare it with an intuitive random strategy (RD) and the two variants of 
selective broadcasting protocol~\cite{agrawal}, i.e. selective broadcasting strategy (SB)~\cite{agrawal} without any 
centralized authority, and selective broadcasting with centralized authority
(CA). 

In RD strategy, channels are randomly selected to be used by CR nodes for transmission and/or
overhearing, i.e. without any consideration to the ongoing PR and CR activity over these channels.
In selective broadcasting SB, each CR node selects a minimum set of channels i.e. Essential Channel
Set (ECS) for transmission, that covers all its geographic neighbors, without considering the PR occupancy.
In our simulations, we consider an implementation of SB with a single transceiver. Thus, transmissions over multiple
channels in the presence of single transceiver is done sequentially with incurred delay, i.e with a round robin process
over the channels of the ECS.  
Regarding message reception, each neighbor sequentially overhears on the channels present in the ECS list. 
Clearly, selecting channels from the ECS one after the other for overhearing reduces the probability of reception 
on each of them.
Selective broadcasting with centralized authority CA, i.e. the third algorithm we compare SURF with, works on the same principle
for transmission as SB, except that each neighbor node simultaneously overhears on all the channels present in the ECS
list.

It is worth noting that selective broadcasting with centralized authority (CA) can be used as a theoretical upper bound 
in message dissemination comparison; since it maximizes the number of receptions by performing overhearing over multiple 
channels, simultaneously. The main difference between SB, CA, RD, and SURF is the number of transmissions generated by 
the first two strategies: 2.5 times more than RD and SURF. Additionally, multiple transmissions of the same message 
over multiple channels may cause multiple receptions of the same message at neighbor nodes, decreasing the
transmission opportunity perceived by CRs. Otherwise, in SURF, nodes switch to a single 
channel based on its occupancy and receivers availability, being no multiple transmissions performed which results in less 
message overhead. Therefore, SURF has an added advantage in this case, as there is no need for {\it a central entity} or any 
other control message to switch overhearing nodes to the same channel on which the neighboring node is transmitting.

We assume that the spectrum opportunity map of PR is available for cognitive radios.
We further consider in our simulations that PR nodes
over every channel switch evenly between ON/OFF states with probability in range [20\%-80\%].
At each CR transmission, the PR
occupancy per channel $i$, ($PR_o^{(i)}$), is calculated according to the number of PR nodes
provided by the opportunity map. Additionally, each CR node locally computes the CR occupancy ($CR_o^{(i)}$)
and the availability level ($P_w^{(i)}$) of each channel $i$. The channel with the highest weight is then
selected for transmission and/or overhearing (cf. Section~\ref{sec:proposal}). The message dissemination
phase then starts, in which a randomly selected CR node disseminates the message on the selected channel by
setting the {\it TTL}. CR neighbor nodes that are on the same selected channel will overhear the message,
decrease {\it TTL}, redo the spectrum sensing, select the best available channel, and disseminate the message
to the next-hop neighbors until {\it TTL=0}.



\subsection{Blocking Ratio}
\label{tunning}

We say that a message is blocked if it is lost because no CR is overhearing over the same channel (while the TTL value is still \textgreater 0).
The {\it Blocking Ratio} is then defined as the number of blocked messages over the total number of sent packets.

\begin{figure}[h]
   \begin{center}
   \includegraphics[width=10cm,height=7cm]{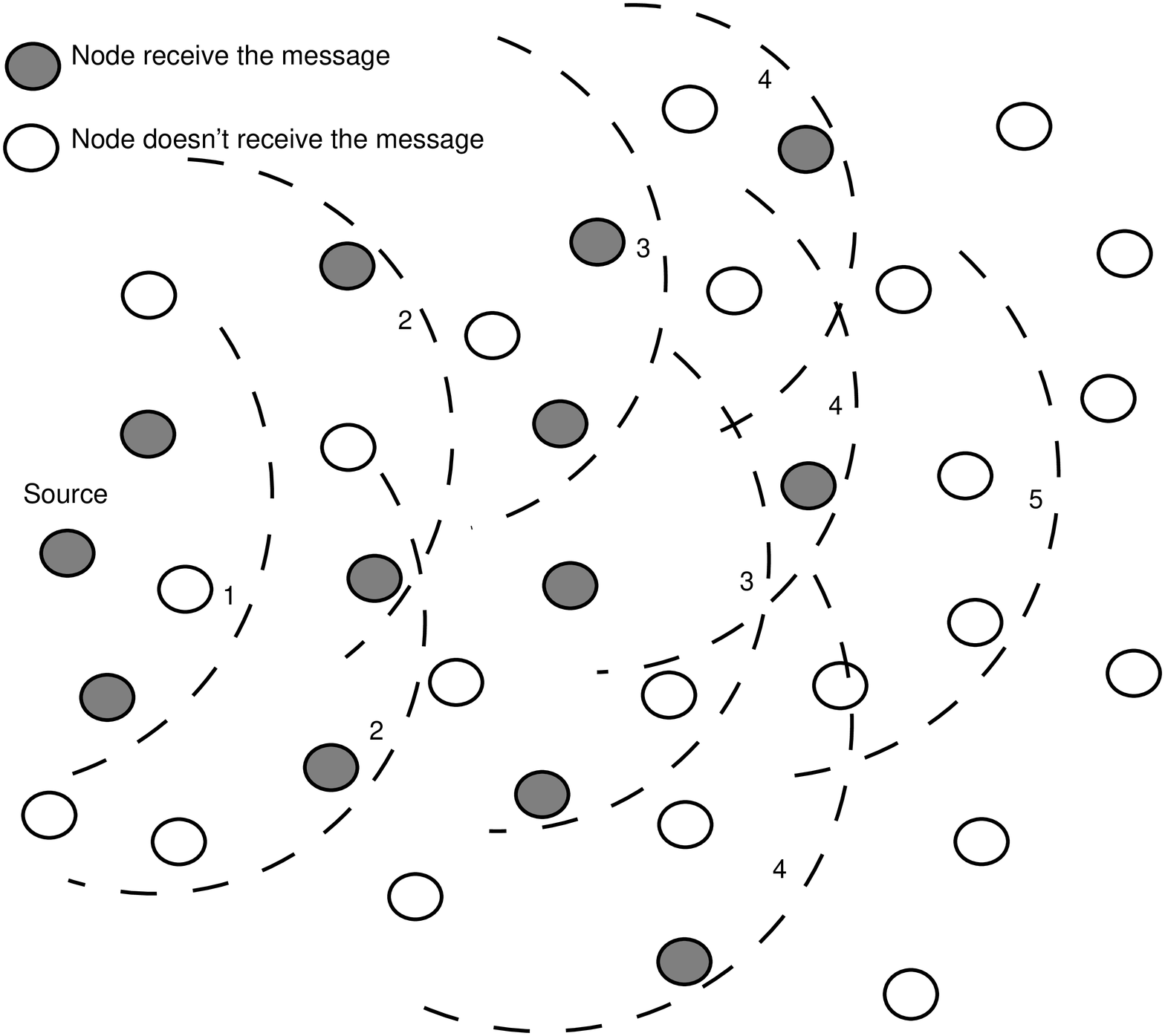}\vspace{-0.1cm}
 \caption{Message blocked after $TTL=4$ in a multi-hop CRN.}
   \label{fig17}
\end{center}
\end{figure}

For effective data dissemination, not only the sender should select the best channel but also the receiver should be tuned
to the right channel (selected by the sender) in order to receive the sent information. Thus, a good channel selection strategy
is the one that tunes both the sender and the receiver to the right channel in the multi-hop context.
Fig.~\ref{fig17} shows, for example, that if receivers nodes are not tuned to the right channel, the dissemination
will be stopped before reaching the highest distant nodes in the network.

\begin{figure}[h]
   \begin{center}
   \includegraphics[width=9cm, height=6cm]{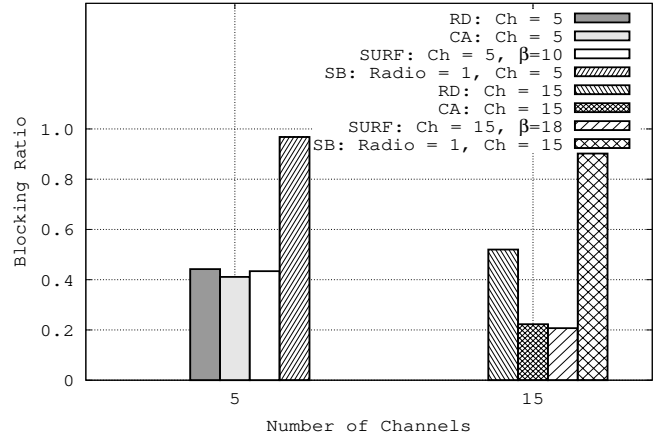} \vspace{-0.1cm}
 \caption{Blocking Ratio in a CRN with 70 CR nodes, for varying number of channels.}
   \label{fig7}
\end{center}
\end{figure}

In Fig.~\ref{fig7}, we compare the blocking ratio obtained with the four strategies detailed above for multi-hop CRN.
Over 1000 sent packets, we compute the blocking ratio caused by receiver not overhearing the appropriate channel.

In a network where 15 channels are available for CR use ($Ch=15$),
RD and SB strategies have higher blocking ratio than SURF and CA.
Such results are highly predictable due to (1) the naive selection approach of the random strategy and (2)
the availability of single transceiver and the lack of any central entity of the SB strategy. In particular,
in SB, the probability of a node overhearing the same channel used for transmission is $\frac{1}{|ECS|}$.

In fact, in both cases when $Ch=5$ and $Ch=15$, the blocking probability of SB is higher than RD.
Practically, the round robin process used at the transmitter {\it and} at the receiver, requires that both of them are
tuned to the same frequency in the same time to correctly receive sent messages. This strict synchronization process is hard to achieve
first because the sender/receiver do not necessarily have the same set of ECS on which they sequentially transmit/receive, and second, a light dephasing on the transmitter or the receiver side may yield to errors in messages reception. Additionally, since SB try to use channels with higher number of neighbors, there is also a chance of
having higher contention and consequently, higher loss ratio than in a random selected channel. The same happens
for CA, where the higher number of CR neighbors increases the contention in each channel, increasing then the number
of blocked messages.

Moreover, the blocking ratio of RD increases with the increase in the total number of channels.
This is because less nodes overhear on the same channel selected by the transmitting node, since CRs are spread
over different channels. On the contrary, an increase in the number of channels has a minor effect on SB.
This is because the increase of total number of channels results in lower sizes of ECS, composed by channels that can
potentially reach more neighbors. Therefore, more nodes have the probability to overhear over the same channel.

SURF has lower blocking probability because our decentralized channel selection makes more nodes overhear over the same channel.
In fact, this happens since during channel selection SURF considers both the PR occupancy and number of CR neighbor receivers.
More surprisingly, SURF has a decrease in blocking ratio as the number of channels increases. This is
mainly due to the fact of having nodes selecting best channels for transmission and reception.

\subsection{Reliability in Data Dissemination}
\label{dd}

To assess the performance of SURF with RD, SB, and CA in term of reliable data dissemination,
two performance metrics are evaluated with different total number of channels:
(i) the average delivery ratio, which is the ratio of packet received by a
particular CR node over total packets sent in the network and (ii) the average number of accumulative CR receivers at
each transmission, until TTL=0. Recall that higher number of channels yields to lower PR occupancy.
In addition, it is worth mentioning here that even the centralized approach CA could not get a 100\% of data
dissemination because of the performed randomly assignment of $Acs$ set to CR nodes. This may generate topology disconnections
caused by physical close nodes being assigned to disjoint channels. In this way, as previously stated, we consider
the CA approach gets the theoretical upper bound results in terms of message dissemination.

\begin{figure*}[!ht]
\begin{center}
\subfigure[Accumulative receivers]{
\includegraphics[width=8.5cm]{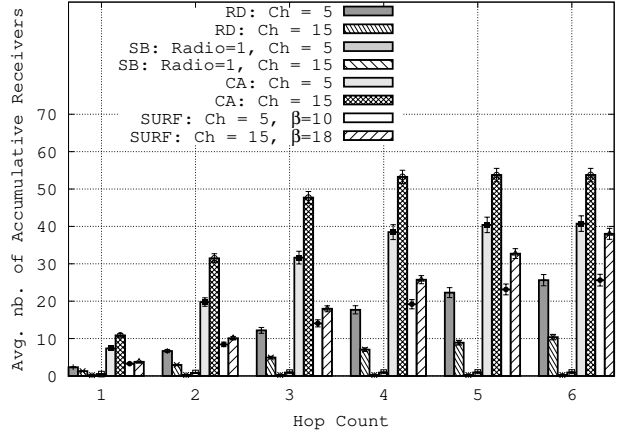}
\label{fig9}
}
\subfigure[Delivery ratio]{
\includegraphics[width=8.5cm]{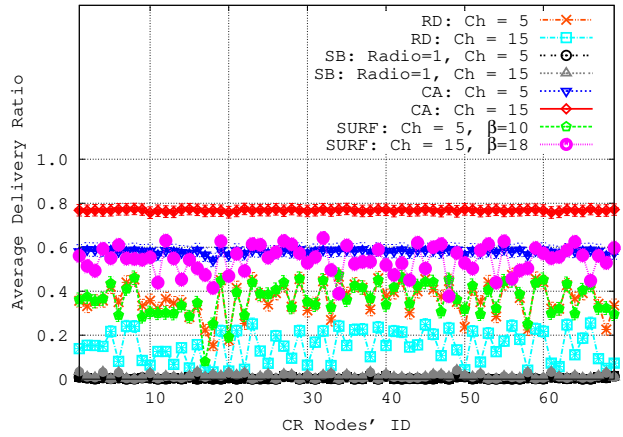}
\label{fig10}
}
\caption{Average number of accumulative receivers per hop and average delivery ratio in a 70-node CRN,
       for random (RD), selective broadcasting (SB), centralized approach (CA), and our strategy (SURF).}
\label{receivers_ratio}
\end{center}

\end{figure*}





Fig.~\ref{fig9} compares the number of accumulative CR receivers at each hop of communication until TTL=0,
for the four strategies. When $Ch=15$, SURF allows the message dissemination to 55\% of nodes in the network (i.e. 38
out of 70 CR nodes), while CA allows 78\% (i.e. 54 over 70 CR nodes).
Additionally, due to its central control and multiple transmissions, the CA strategy reaches this upper bound of receivers
percentage at the TTL=4.
It can be clearly seen that SURF outperforms RD and SB and compared to CA, only provides a decrease of 25\% in performance. The gain
achieved with CA is at the price of more transmissions, more energy consumption, and more expensive and sophisticated devices.

Fig.~\ref{fig10} compares delivery ratio of RD, SB, CA and SURF, as a function of the CR nodes' ID. SURF outperforms RD
and SB in terms of delivery ratio, when number of channels are high. Compared to the CA strategy, SURF has only
20\% of performance reduction. In particular, for {\it Ch=5} and {\it Ch=15}, SURF guarantees the delivery of
approximately 60\% of messages (with a single transmission), contrarily to less
than 20\% for the RD and SB strategies (with single and multiple transmissions, respectively) and
80\% for the CA strategy (with multiple transmissions).

\section{Advantages of SURF}
\label{adv}
SURF, by exploiting information regarding PR and CR occupancy, brings several advantages. Some of them are highlighted below:
\begin{itemize}
\item Less interference with PR nodes: Through our channel selection strategy, CR nodes are bound to select those
channels which are less utilized by PR nodes. Therefore they cause less interference to PR nodes thus, satisfying
the major constraint of CRN.
\item Autonomy and decentralization: CR nodes make local and distributed decision for channel selection.
Therefore SURF makes CR nodes autonomous in their channel selection decision.
\item No control messages exchange: Implementing the same strategy at the sender and receiver helps both of them tune to the
appropriate channel for undergoing transmissions or reception without the need of any prior information exchange or
synchronization.
\item Less overhead: As SURF is based on single transmission, it generates less overhead compared to
channel selection strategies based on multiple transmissions (e.g. \cite{agrawal}).
\item Practical feasibility and low cost: A key characteristic of our channel selection strategy is that it assumes the availability
of single transceiver, which is used for both transmission and/or overhearing. It reduces thus, the operational cost
of the network.
Besides reducing transmissions overhead, transmitting over a single channel
cuts down energy consumption and increases the battery lifetime of CRs.
\item Information relaying: Another key advantage of our multi-hop channel selection strategy is that the same
strategy can be reused and reconfigured to relay information from one or more users to others receiving users located in
various locations in the network.
\item Network coverage improvement: Improved network performance in terms of CR network' perspective is achieved
by making CR nodes to switch to highly reliable channels, which as a consequence, increase network coverage.
\end{itemize}

\section{Related Work}
\label{related}

Recently, a lot of work has been carried out for dynamic channel management in cognitive radio 
networks~\cite{cao,pomdp,niyato,rahul}. However, all these approaches focuses on single-hop 
cognitive radio networks and either requires the presence of any central entity or coordination with primary
radio nodes in their channel selection decision. For instance,~\cite{cao} proposed an efficient spectrum 
allocation architecture that adapts to dynamic traffic demands but they considered a single-hop scenario of 
Access Points (APs) in Wi-Fi networks.~\cite{mubashir} proposed a channel selection strategy based on the primary 
user's occupancy but specifically designed for single-hop architecture. 

In this paper, we focus on channel selection in the context of multi-hop cognitive radio ad hoc networks, 
where no cooperation or feedback is expected from primary nodes and the network operates in the absence of any 
centralized authority. In addition, an adaptive channel selection strategy is required at both the sender and 
receiver node, so that the receiver node tuned to the right channel to receive sent information. Moreover, 
the holding time and the granularity of wireless spectrum bands also affects on multi-hop CR communications~\cite{hicham1}. 
All these factors makes channel selection in these networks extremely challenging, having very few works been done 
so far~\cite{agrawal,distributed}. In~\cite{distributed}, the authors proposed a dynamic resource management scheme 
for multi-hop cognitive radio networks. But their approach is based on periodic 
control information exchange among nodes, which is not the case in SURF.


In selective broadcasting (SB)~\cite{agrawal}, each cognitive node selects a minimum set of channels (ECS) covering 
all of its geographic neighbors to disseminate data 
in multi-hop cognitive radio networks. There are however, several challenges in the practicality of SB. 
Indeed, from the communication perspective, simultaneous transmission over a ECS requires more than one 
transceiver, which means having bigger and more complex 
devices, as it is done for military applications~\cite{ossama}.

%


\section{Conclusion and Future Work}
\label{sec:conclusion} 
In this paper we have proposed SURF, a channel selection strategy for reliable contention-aware data 
dissemination in multi-hop cognitive radio network. 
SURF selects a single channel for every message transmission allowing the best opportunities for CR-to-CR communications due to 
(1) low primary radio nodes activities, and (2) limited contention of cognitive radio nodes acceding that channel. 
The result is a connected network with limited contention, where reliable communication can take place. 
SURF strategy is simple, completely decentralized, and is based on practical assumptions. Thus, it can be applied 
today to many possible cognitive radio networks deployments to deliver emergency alerts and advertisement messages.
We have demonstrated through simulations the performance of our channel selection strategy when 
compared with three other related approaches. Simulation results confirmed that our approach 
is effective in selecting the best channels for efficient and reliable multi-hop data dissemination. 

As plan of our future work, we intend to investigate the time needed to disseminate messages in the network. This 
delay will surely depend on the size of the network and on the number of available channels at CR nodes. 
Besides, empowering SURF with channel history components that assist the channel selection process and mainly
that tie-breaks when two or more channels have the same weight is also a future research direction.  


\bibliographystyle{plain}
\bibliography{WoWMoM}

%

\tableofcontents
\end{document}